\documentclass[preprint,showpacs,preprintnumbers,amsmath,amssymb]{revtex4}
\usepackage{graphicx}

\begin{document}

\thispagestyle{empty}

\title{Comparison of the experimental data for the
Casimir pressure with the Lifshitz theory at zero
temperature}

\author{ B.~Geyer,${}^{1}$
G.~L.~Klimchitskaya,${}^{1,2}$  and
V.~M.~Mostepanenko${}^{1,3}$}

\affiliation{
${}^1$Institute for Theoretical
Physics, Leipzig University, Postfach 100920,
D-04009, Leipzig, Germany \\
${}^2${North-West Technical University,
Millionnaya Street 5, St.Petersburg,
191065, Russia}\\
${}^3${Noncommercial Partnership ``Scientific Instruments'',
Tverskaya Street 11, Moscow,
103905, Russia}
}

\begin{abstract}
We perform detailed comparison of the experimental data of the
experiment on the determination of the Casimir pressure
between two parallel Au plates with the theoretical values
computed using the Lifshitz formula at zero temperature.
Computations are done using the optical data for the complex
index of refraction of Au extrapolated to low frequencies by means
of the Drude model with both most often used and other suggested
Drude parameters. It is shown that the experimental data exclude
the Lifshitz formula at zero temperature at a 70\% confidence
level if the Drude model with most often used values of the
parameters is employed. If other values of the Drude parameters are
used, the Lifshitz formula at zero frequency is experimentally
excluded at a 95\% confidence level. The Lifshitz formula at zero
temperature combined with the generalized plasma-like model with
most often used value of the plasma frequency is shown to be
experimentally consistent. We propose a decisive experiment
which will shed additional light on the role of relaxation
properties of conduction electrons in the Casimir effect.
\end{abstract}
\maketitle

\section{Introduction}

There is an increasing interest in the Casimir effect \cite{1}
in the recent literature connected with numerous
multidisciplinary applications in both fundamental and
applied science (see monograph \cite{2} for a modern overview
of the subject). The Casimir force acting between two closely
spaced uncharged material bodies is connected with the
existence of zero-point and thermal fluctuations of the
electromagnetic field. Keeping in mind that in some sense the
vacuum is the most fundamental quantum state, the role of the
Casimir force in many diverse areas ranging from elementary
particles and gravitation to atomic physics, condensed matter
physics and nanotechnology becomes clear. Over a long
period of time the experimental investigation of the Casimir
effect has progressed only slowly because the related forces and
energies are very small and their observation requires
special conditions which are hard to achieve. During the last
12 years, however, about 25 experiments measuring the
Casimir force have been performed using new possibilities
suggested by modern laboratory techniques (the review \cite{3}
describes all
recent experiments and related theory).

It is well known that the most fundamental theoretical
description of the van der Waals and Casimir forces acting
between two material bodies is given by the Lifshitz theory
\cite{4,5,6} (recall that in this case the Casimir force
is nothing but the retarded van der Waals force).
The Lifshitz theory was originally formulated for two
semispaces separated by a gap. Recently far-reaching
generalizations of the Lifshitz theory have been proposed
allowing calculation of the Casimir force between arbitrarily
shaped bodies (see, for instance,
Refs.~\cite{2,7,8,9,10,11,12,13,14,15}).
In the Lifshitz theory and its generalizations the Casimir
energy and force are expressed in terms of reflection
amplitudes describing reflection of the electromagnetic
oscillations on the boundary surfaces. If the spatial dispersion
of material bodies can be neglected, the reflection amplitudes
in turn are expressed using the dielectric permittivity
depending on the frequency $\omega$. This is in fact the basic
quantity in the Lifshitz theory which should be known to
calculate the Casimir energy (free energy) and the Casimir
force.

The comparison of the experimental data with the
Lifshitz theory at nonzero temperature  has revealed
a puzzle which remains unresolved to the present day \cite{3}. Some
authors \cite{16,17,18} (see complete set of
references in review \cite{3}) consider most natural
the suggestion to substitute in the Lifshitz formula the
full dielectric permittivity taking into account all physical
processes really occurring at the corresponding frequencies.
This leads to the use of dielectric permittivity containing
the first order pole at $\omega=0$ due to the role of
conduction electrons.
The respective behavior of the dielectric permittivity at
low frequencies is usually described by the Drude model.
However, the experimental data of several experiments
performed at nonzero temperature with metallic \cite{19,20,21,22},
semiconductor \cite{23,24} and dielectric \cite{25,26} test
bodies are inconsistent with the theoretical predictions of the
Lifshitz theory obtained using this suggestion.
Different attempts on how to resolve the
puzzle of the thermal Casimir force, including the question
of reliability of the data, were discussed \cite{27}
(see also in Secs.~II and III below).

The use of
the Lifshitz formula at zero temperature $T=0$ is
one of the widespread approaches to the comparison of
measurement data with theory in Casimir physics \cite{28,29,30,31}.
This is usually justified by stating
that at small separations between the test bodies the corrections
to the Casimir force due to nonzero temperature are
insignificant. It should be noted that the role of nonzero
temperature in the Lifshitz formula is twofold: in the
discreteness of the imaginary frequencies at $T\neq 0$ and in
the explicit dependence of the dielectric permittivity on $T$.
What is commonly referred to as ``the Lifshitz formula at
zero temperature'', takes into account only the first factor.
This means that
the integration with respect to a continuous frequency is
performed instead of summation over the discrete Matsubara
frequencies. In so doing the second factor is disregarded, i.e.,
the dielectric permittivity, as measured at  room temperature,
is preserved. The ``hybrid'' character of such kind of
``zero-temperature'' formula was investigated \cite{32}.
Specifically, it was shown that respective ``zero-temperature''
Casimir energy, even at short separations, can deviate from
the Casimir free energy computed at room temperature by
several percents. Nevertheless it is rather common to believe \cite{33}
that the Lifshitz formula at zero temperature ``gives a
dominant contribution at small separations ($<1\,\mu$m at room
temperature) between the bodies and was readily confirmed
experimentally with good accuracy\ldots''.

In this paper we perform the detailed comparison of the
 experimental data on an indirect measurement of the
Casimir pressure between two parallel plates \cite{21,22} with
the Lifshitz formula for the Casimir pressure at zero temperature.
This comparison is performed using the tabulated \cite{34} optical data
for Au  extrapolated  to low frequencies
by means of the Drude model with most often used Drude
parameters  \cite{39} and
with other Drude parameters, as suggested, e.g., in
Ref.~\cite{35}. Our comparison shows that the experimental data
exclude the Lifshitz formula at zero temperature, which uses the
tabulated optical data extrapolated to low frequencies with the help
of most often used Drude parameters,
at a 70\% confidence level over a wide
separation region. The zero-temperature Lifshitz formula using
other Drude parameters is excluded by
the data at a 95\% confidence level. According to our results,
if the experiment is performed at room temperature, the Lifshitz
formula also at room  temperature should be used to make
a comparison between the data and the theory. We discuss a recent
suggestion \cite{37} on how to avoid the use of ad hoc
extrapolations of the optical data outside the frequency region
where they were measured.
This suggestion is based on some properties of analytic functions
but meets difficulties in practical realization.
We also consider the Lifshitz formula at zero temperature using
the generalized plasma-like model \cite{2,3,36}, which disregards
relaxation properties of conduction electrons, and compare the
computational results with the same experimental data.
It was shown that in this case the data are consistent with
the theory employing the most often used value of the plasma frequency.
This is explained by the fact that the generalized
plasma-like model at separations below $1\,\mu$m leads to
approximately the same results irrespective of whether the
Lifshitz formula at zero or nonzero temperature is used.
We also propose a new experiment which can shed additional light
on the role of relaxation properties of conduction electrons
in the Casimir effect.

The paper is organized as follows. In Sec.~II we compare the
experimental data \cite{21,22} with the
zero-temperature Lifshitz formula which utilizes the extrapolation
of the optical data by the Drude model with most often used
parameters. In Sec.~III the same data are compared with the same
formula, but with other Drude parameters.
The possibility on how to determine the dielectric permittivity
along the imaginary frequency axis using only the measured optical
data is discussed in Sec.~IV.
Section V is devoted to the comparison of the experimental
data with the Lifshitz formula at zero temperature combined
with the generalized plasma-like model. In Sec.~VI the reader will
find our conclusions and discussion including the proposal
of new decisive experiment.

\section{Comparison of the experimental data with theory
{\protect \\} using
conventional extrapolation of the optical data by the Drude
model}

Here and below we use the experimental data of the
experiment on an indirect measurement of the Casimir pressure
between two parallel plates by means of micromachined
oscillator \cite{21,22}. This experiment used the configuration
of a Au-coated sphere of $150\,\mu$m radius above a Au-coated
plate that could rotate about the rotation axis. During the
measurements, the separation between the sphere and the plate
was varied harmonically at the resonant frequency of the oscillator.
The immediately measured quantity was the shift in this frequency
due to the Casimir force acting between the sphere and the plate.
Using the proximity force approximation \cite{2,3,38},
the shift of the resonant frequency of the oscillator was
recalculated into the equivalent Casimir pressure in the
configuration of two parallel plates made of Au. The pressure was
determined as a function of separation over the separation
region from 160 to 750\,nm.
The absolute error in the measurement of separation
distances $a$ was determined to be $\Delta a=0.6\,$nm
at a 95\% confidence level. The absolute error in the
determination of the Casimir pressure was also determined
at a 95\% confidence level and found to be
separation-dependent.
The respective relative error increases from approximately 0.2\%
at $a=160\,$nm to 9\% at $a=750\,$nm.
It was shown that in this experiment
the total experimental error is completely determined
by the systematic error leaving the random error negligibly
small, as it should be in precise experiments of
metrological quality (details of the measurements,
calculations and error analysis can be found in
Refs.~\cite{2,3,21,22}).

According to our aim, the theoretical Casimir pressure
between smooth parallel plates is computed using the
Lifshitz formula at zero temperature
\begin{equation}
P(a)=-\frac{\hbar}{2\pi^2}\int_{0}^{\infty}d\xi
\int_{0}^{\infty}k_{\bot}dk_{\bot}q\sum\limits_{\alpha}
\left[\frac{e^{2aq}}{r_{\alpha}^2(i\xi,k_{\bot})}-1
\right]^{-1}.
\label{eq1}
\end{equation}
\noindent
Here, $\omega=i\xi$, $k_{\bot}$ is the projection of the wave
vector onto the plane of the plates,
 $\alpha$ denotes the
transverse magnetic (TM) and transverse electric (TE)
polarizations of the electromagnetic field,
and $q=(k_{\bot}^2+\xi^2/c^2)^{1/2}$. The respective
reflection coefficients are
\begin{eqnarray}
&&
r_{\rm TM}(i\xi,k_{\bot})=
\frac{\varepsilon(i\xi)q-k}{\varepsilon(i\xi)q+k},
\nonumber \\
&&
r_{\rm TE}(i\xi,k_{\bot})=\frac{q-k}{q+k},
\label{eq2}
\end{eqnarray}
\noindent
where
$k=[k_{\bot}^2+\varepsilon(i\xi)\xi^2/c^2]^{1/2}$ and
$\varepsilon(i\xi)$ is the dielectric permittivity of the
material calculated along the imaginary frequency axis.

We have performed computations by Eqs.~(\ref{eq1}) and
(\ref{eq2}) within the experimental separation region from
160 to 750\,nm. The dielectric permittivity of Au along the
imaginary frequency axis was found by means of the
Kramers-Kronig relation \cite{2} which assumes that
$\varepsilon(\omega)$ is regular or has a first-order pole
at $\omega=0$:
\begin{equation}
\varepsilon(i\xi)=1+\frac{2}{\pi}\int_{0}^{\infty}
\frac{\omega\,{\rm Im}\,\varepsilon(\omega)}{\xi^2+\omega^2}
\,d\omega.
\label{eq3}
\end{equation}
\noindent
This was done using the tabulated optical data \cite{34}
for the imaginary part of the dielectric permittivity,
${\rm Im}\,\varepsilon(\omega)$, measured in the frequency
region from 0.125 to $10^4\,$eV and extrapolated to lower
frequencies by means of the Drude model
\begin{equation}
\varepsilon(i\xi)=1+\frac{\omega_p^2}{\xi(\xi+\gamma)}.
\label{eq4}
\end{equation}
\noindent
The values of the plasma frequency $\omega_p$ and relaxation
parameter $\gamma$ were determined \cite{22} from
the measurements of resisitivity of the used Au films as a
function of temperature. These values ($\omega_p=8.9\,$eV and
$\gamma=0.0357\,$eV) are very close to the  values \cite{34,39}
$\omega_p=9.0\,$eV and $\gamma=0.035\,$eV
most often used
in numerous calculations of the Casimir force between Au
surfaces made by different authors (see review \cite{2,3}).
It was not needed to use any extrapolation of the optical data
to higher frequencies.

The influence of surface roughness was taken into account in a
nonmultiplicative way using the method of geometrical averaging
\cite{2,3,20,22}. According to this method the theoretical Casimir
pressures between the rough plates were calculated as
\begin{equation}
P^{\rm theor}(a)=\sum\limits_{i=1}^{N_p}\sum\limits_{j=1}^{N_s}
v_j^{(s)}v_i^{(p)}P(a+H_s+H_p-h_j^{(s)}-h_i^{(p)}).
 \label{eq5}
\end{equation}
\noindent
Here, the surface topography of the plate (sphere) is approximately
characterized  by $N_p$ ($N_s$) pairs $(v_i^{(p)},h_i^{(p)})$
[$(v_j^{(s)},h_j^{(s)})$], where $v_i^{(p)}$ ($v_j^{(s)}$) is the
fraction of the surface area with height $h_i^{(p)}$ ($h_j^{(s)}$).
These data obtained \cite{20} from atomic force microscope scans
allow one to determine the zero-roughness levels $H_p$ ($H_s$) relative
to which the mean values of roughness profiles are equal to zero
\begin{equation}
\sum\limits_{i=1}^{N_p}(H_p-h_i^{(p)})v_i^{(p)}=0,
\quad
\sum\limits_{j=1}^{N_s}(H_s-h_j^{(s)})v_j^{(s)}=0.
 \label{eq6}
\end{equation}

Now we compare the computational results for the Casimir pressure at
zero temperature, $P^{\rm theor}(a)$, with the experimental
data \cite{22}. In Fig.~1(a) the computational results at separations
 from 350 to 400\,nm are shown as
bands in between the two solid lines. The width of a theoretical band
is determined by the total theoretical error of about 0.5\% found
at a 95\% confidence level \cite{20}. This includes errors due to
uncertainty of the optical data \cite{34}, contribution
of patch potentials, and diffraction-type contribution to the effect
of surface roughness which is not taken into account by the method
of geometrical averaging (see details \cite{20}).
The contribution of patch potentials due to grains of polycrystal
Au films was estimated \cite{20}.
At the shortest separation $a=160\,$nm it was shown to be only
0.037\% of the Casimir pressure, and further decreases with increasing
$a$. The diffraction-type contribution to the effect of surface roughness
is less \cite{20} than 0.04\% at $a=300\,$nm. Although it increases
with increasing $a$, the total effect of surface roughness becomes
negligibly small \cite{20} at $a>300\,$nm.
The mean experimental Casimir pressures,
$\bar{P}^{\rm expt}(a)$, are shown as crosses whose arms are also
determined at a 95\% confidence level. All details on the measurement
procedures used for measuring both the pressures and absolute separations
and determination of experimental errors are
presented \cite{20,21,22}.
Specifically, the total experimental error
of pressure measurements (which is mostly determined by the systematic
error) includes the error due to use of the proximity force
approximation to convert the data for the frequency shift into the
data for the Casimir pressure. As can be seen in Fig.~1(a), all
experimental crosses lie outside of the theoretical bands, but some of
the arms of the crosses touch the border lines of these bands. Thus,
in the strict sense one cannot claim that the theoretical description
using
the Lifshitz formula at zero temperature
combined with the Drude model is excluded by the data at
a 95\% confidence level.

Let us now perform the comparison of the experimental data with the
zero-temperature theoretical results at a lower, 70\%, confidence level.
For this purpose we assume that both the theoretical and experimental
errors are random quantities distributed uniformly (other
hypothesis would lead to smaller errors at a 70\% confidence level
so that our approach is the most conservative \cite{40}).
Now, to obtain the theoretical band and the arms of the experimental crosses
defined at a 70\% confidence level one should divide their widths
in Fig.~1(a) by a factor of 0.95/0.7=1.357. The resulting comparison of
experiment with theory at a 70\% confidence level is presented in
Fig.~1(b) within the separation
region from 350 to 400\,nm. The same notation
as in Fig.~1(a) is used. As can be seen in Fig.~1(b), all experimental crosses
are outside the theoretical band confined between the solid lines.
This means that the Lifshitz theory at zero temperature employing the
optical data extrapolated to zero frequency by means
of the Drude model with most often used parameters
is excluded by the data of the experiment \cite{21,22}
at a 70\% confidence level.

The obtained conclusion can be confirmed using another method for
the comparison between experiment and theory based on the consideration of
a confidence interval for the differences of theoretical and mean
experimental Casimir pressures
$P^{\rm theor}(a)-P^{\rm expt}(a)$ calculated at all experimental
separations (see Refs.~\cite{2,3,20,22,41}).
At a given confidence level, such intervals
$[-\Xi(a),\Xi(a)]$ are different at different
separations.  The confidence interval
$[-\Xi_{0.95}(a),\Xi_{0.95}(a)]$ was found \cite{22}
at a 95\% confidence level for the
differences between predictions of the room-temperature Lifshitz
theory and the experimental data \cite{21,22}. It was obtained as a
combination of the total theoretical and total experimental errors.
Keeping in mind that the total theoretical error is almost independent
of the temperature and model of the dielectric
permittivity \cite{2,3,20,22},
we can use the same confidence interval for the
comparison between the experimental data and the theoretical results
obtained using the zero-temperature Lifshitz formula.

In Fig.~2 the borders of the confidence intervals
$[-\Xi_{0.95}(a),\Xi_{0.95}(a)]$ generate the two solid lines.
The differences between the theoretical Casimir pressures computed using the
zero-temperature Lifshitz formula and the tabulated optical data
extrapolated to zero frequency are indicated as dots.
As can be seen in Fig.~2, within the separation intervals
$a<310\,$nm and $a>460\,$nm all dots are inside the confidence intervals,
i.e., the theoretical approach used is formally consistent with the
data. However, within the interval from 310 to 460\,nm many dots are
on the border of the confidence intervals or even outside of them.
This casts some doubts on the consistency of the used theoretical
approach with the data and calls for the consideration of confidence
intervals at lower, 70\%, confidence level. For this purpose we have
investigated the distribution law of the random quantity
$P^{\rm theor}(a)-P^{\rm expt}(a)$ near its mean value over the entire
measurement range. We have found that to sufficient accuracy this
distribution is normal. Thus, the desired half-width of the confidence
interval at a 70\% confidence level can be determined from the
equality $\Xi_{0.95}(a)/\Xi_{0.7}(a)=2$.
The obtained borders of the confidence intervals
$[-\Xi_{0.7}(a),\Xi_{0.7}(a)]$ generate the two dashed lines in Fig.~2.
As can be seen in this figure, over a wide separation region from 230
to 520\,nm all dots lie outside the confidence intervals. This means that
theoretical approach employing the zero-temperature Lifshitz formula and
extrapolation of the optical data to zero frequency by means
of the Drude model with most often used parameters
is experimentally excluded at a 70\% confidence level.

We complete this section with a brief discussion of the reliability of
used experimental results and their comparison with theory.
As often underlined in the literature (see, e.g., Ref.~\cite{35}),
in all performed experiments both the optical data for metallic films
and the values of the Drude parameters used to extrapolate
data to lower frequencies were not measured but taken from
handbook~\cite{34}. It was shown \cite{42}, however, that the variation of the
optical data for Au
for different samples may influence on the Casimir force
on the level of 5\%.
The question arises whether or not this could influence
the validity of the above conclusion that the Lifshitz
theory at $T=0$ combined with the Drude model
is experimentally excluded. First of all we recall that
in the experiment \cite{21,22} the values of the Drude parameters
were determined from the measurement of resistivity of the used films
as a function of $T$. Recently one more measurement
of the Casimir pressure by means of a micromachined
oscillator was performed
on a Au electroplated sample where
 the optical data were obtained \cite{43}
by ellipsometry in the
frequency region from 1.50 to 6.25\,eV. It was shown that the
experimental results for the Casimir pressure in Refs.~\cite{21,22} and
\cite{43} are virtually undistinguishable. The differences between
the measured and tabulated optical data are very small and do not affect the
computational results for the Casimir pressure \cite{43}.

Another point that could influence the experimental results is a possible
uncertainty in the electrostatic calibrations used to determine the
absolute separation between the test bodies, sphere radius and some other
parameters. Thus, an
anomalous distance dependence of the electric force acting between an
Au-coated plate and an Au-coated sphere of 30\,mm radius
was observed \cite{44}.
The respective contact potential was found to be dependent on separation.
In experiments using micromachined oscillator the standard force-distance
dependence was observed
and the contact potential was measured to be constant  \cite{20,21,22,45,46}.
The observed anomalous distance dependence \cite{44} might be
explained \cite{45}
by deviations of the mechanically polished and ground surface
from a perfect spherical shape  for lenses
of centimeter-size radius.
Recently the same conclusion has been made \cite{47}
for cylindrical surfaces of large radii.
Notice that unexpected features of the electrostatic calibrations in the
measurements of the Casimir force
between metal bodies
were also reported by some other
authors \cite{49,50}.
We emphasize, however, that in all
these cases either spheres of centimeter-size radius have been used or
the experiments were performed in an ambient environment (here we do
not discuss the case of semiconductor test bodies where the effect
of space-charge layer should be taken into account at short
separations \cite{2,3}).

The authors \cite{44,50} relate the anomalies in electrostatic
calibrations observed in their experiments with large spherical lenses
to possible influence of patch potentials. According to Ref.~\cite{50},
for small patches with effective area $S_p\ll S_{\rm eff}=2\pi Ra$,
where $R$ is the sphere (lens) radius, the additional electric force
arising due to the existence of patches exponentially vanishes with the
increase of separation. In the opposite case of large patches satisfying the
condition $S_p>S_{\rm eff}$,  the possibility
of large additional electric force arises \cite{50}.
The respective potential which minimizes
the total electric force acting between a sphere of centimeter-size
radius and a plate after some voltage is applied, becomes
separation-dependent. As was mentioned above, in the
experiments \cite{20,21,22}
the contact potential does not depend on separation. The investigation of
the surfaces of a sphere and a plate by means of an atomic force
microscope \cite{20} demonstrated that the maximum diameter of grains is
equal to $D=300\,$nm. We emphasize that in the experiments \cite{20,21,22}
the contact potential does not depend on separation and, thus,
patches are caused solely by the grain structure of the sphere and plate
surfaces which are spherical and plain, respectively, otherwise.
 Taking into
account that for the sphere used in the
experiment \cite{21,22} it holds
\begin{equation}
S_p=\frac{\pi D^2}{4}=0.07\,\mu\mbox{m}^2\ll
2\pi Ra=150.72\,\mu\mbox{m}^2,
\label{eq7}
\end{equation}
\noindent
one arrives at the conclusion that only the small patches might be of
relevance to the experiments \cite{20,21,22} (we substituted the shortest
separation $a=160\,$nm in this estimation).
The influence of just such patches was
 analyzed \cite{20} on the basis of
the theory developed \cite{51}
and confirmed \cite{50} recently, and their role was shown
to be negligibly small.

\section{Extrapolations of the optical data by the Drude model with
alternative parameters}

Here, we compare the theoretical predictions of the Lifshitz formula
at zero temperature using the Drude model with other
suggested  parameters
with the experimental data \cite{21,22}. The other
Drude parameters were obtained for Au films of different thicknesses
deposited on different substrates, unannealed or annealed after the
deposition \cite{35}. The optical properties of these films were
measured ellipsometrically within the frequency region from 0.0376 to
0.653\,eV and from 0.729 to 8.856\,eV. Note that the lowest
frequency of
the first interval is a factor of 3.3 smaller than the minimum
frequency where optical data are available in handbook \cite{34}.
For the determination of the Drude parameters $\omega_p$
and $\gamma$ the joint fit of both real and imaginary parts of the
dielectric permittivity or the complex index of refraction to the
optical data was performed in the low frequency range. The consistency
of the obtained complex dielectric permittivity with the
Kramers-Kronig relations was verified \cite{35}. This had led to
different sets of the mean Drude parameters for 5 different samples of
Au films varying from $\omega_p^{(1)}=(6.82\pm 0.08)\,$eV,
$\gamma^{(1)}=(40.4\pm 2.1)\,$meV for the first sample to
$\omega_p^{(5)}=(8.38\pm 0.08)\,$eV,
$\gamma^{(5)}=(37.1\pm 1.9)\,$meV for the fifth sample.

Below we perform computations of the Casimir pressure at zero temperature
using Eq.~(\ref{eq1}) and the extrapolation of the optical data to
the low frequencies by the Drude model with the plasma frequency
varying from $\omega_{p,\min}^{(1)}=6.74\,$eV to
$\omega_{p,\max}^{(5)}=8.46\,$eV.
In so doing the dielectric permittivity along the imaginary
frequency is found from Eq.~(\ref{eq3}).
In the frequency region above 0.125\,eV we continue using the optical
data for the imaginary part of dielectric permittivity
from handbook \cite{34}. This is justified as follows.
In the frequency region from 2 to 6\,eV containing the first two
absorption bands of Au the optical data for ${\rm Im}\varepsilon(\omega)$
measured \cite{35} differ from that in handbook \cite{34}.
Specifically, for sample N2 (N3) the maximum of the first absorption
band is a factor of 0.69 (0.85) of the respective handbook
maximum \cite{34}. The maximum of the second absorption
band for sample N2 (N3) is less than the maximum of the second absorption
band \cite{34} by a factor of 0.64 (0.8).
This can be seen in Fig.~4 of Ref.~\cite{35}.
However, when the optical data \cite{35}
in the region of a few eV are replaced with the data of
 handbook \cite{34}, this results in a negligibly small variation on the
Casimir pressure. For sample N2, the magnitude of the Casimir pressure at
separations 160 and 200\,nm is increased by 0.7\% and 0.5\%,
respectively. Similarly, for sample N3, increases of  0.3\% and 0.2\%
in the magnitude of the Casimir pressure  are obtained at 160 and 200\,nm,
respectively, when data \cite{35} are replaced by those
from handbook \cite{34}.
Thus, we can use the optical data \cite{34} for the first
absorption bands in our computations. Moreover,
the use of the optical data \cite{35} instead of the
handbook data \cite{34} would {\it decrease} the magnitude of the
theoretical Casimir pressure and, thus, only {\it increase}
discrepances between the predicted and experimental Casimir pressures
(see Fig.~1). Note also
 that the optical data measured in Ref.~\cite{43}  are in
very good agreement with the handbook data \cite{34} (only 0.05\% and
0.03\% difference of the Casimir pressure at separations of 160 and 200\,nm,
respectively).

The computational results for the Casimir pressure (\ref{eq1})
are presented in Fig.~3
as a band enclosed in between two solid lines in separation regions (a)
from 300 to 350\,nm and (b) from 350 to 400\,nm. The upper solid lines
in figures (a) and (b) are computed with the help of extrapolation of
the optical data by the Drude model with the alternative parameters
$(\omega_{p,\min}^{(1)},\gamma^{(1)})$. The lower solid lines in figures
(a) and (b) are computed with the parameters
$(\omega_{p,\max}^{(5)},\gamma^{(5)})$. The obtained upper (lower)
lines are shifted
upwards (downwards) by 0.5\%  to take
into account the theoretical error discussed in Sec.~II.
The computational results for the samples \cite{35}
N2, N3 and N4 are sandwiched between these solid lines. (Note that
the value of the
relaxation parameter $\gamma$ only slightly influences the computational
results for the Casimir pressure; for example, a shift in the value
of $\gamma^{(1)}$ by 5\% leads to a shift in the value of the Casimir
pressure varying from 0.07\% to 0.1\% when separation varies from 160
to 750\,nm.)
In the same figure the experimental data
are shown as crosses which arms are drawn at a 95\% confidence level.
As can be seen in Fig.~3, the Lifshitz formula at zero temperature
using the Drude model with other
suggested parameters for the extrapolation
of the optical data to lower frequencies is excluded by the experimental
data \cite{21,22} at a 95\% confidence level.

The same conclusion is obtained if one uses the comparison between
experiment and theory in terms of the differences of calculated and
measured Casimir pressures as described in Sec.~II.
In Fig.~4(a) the upper dots show the differences
$P^{\rm theor}(a_i)-\bar{P}^{\rm expt}(a_i)$, where the values
$P^{\rm theor}(a_i)$ are computed as described above using the
extrapolation of the optical data by the Drude model with the parameters
$(\omega_{p,\min}^{(1)},\gamma^{(1)})$. The lower dots use
the Drude extrapolation with the parameters
$(\omega_{p,\max}^{(5)},\gamma^{(5)})$.
The dots related to all other samples are sandwiched between these two sets
presented in Fig.~4(a). The solid lines show the borders of the
confidence interval $[-\Xi_{0.95}(a),\Xi_{0.95}(a)]$ determined at a 95\%
confidence level. As can be seen in Fig.~4(a), the zero-temperature
Lifshitz formula using the optical data and the Drude model with
other suggested parameters is experimentally excluded
at a 95\% confidence level over a wide range
of separations from 160 to 520\,nm.
In Fig.~4(b) the same two sets of differences between the computed and
mean measured Casimir pressures are shown over a narrower separation
region from 500 to 750\,nm. Here, in addition to the solid line indicating
the borders of a 95\% confidence interval, the dashed lines indicate
 the confidence intervals determined at a 70\%
confidence level. As can be concluded from Fig.~4(b), at a 70\%
confidence level the zero-temperature Lifshitz formula using the optical
data and other suggested Drude parameters is experimentally excluded
over an even wider separation range from 160 to 620\,nm.

\section{Suggestion to use a window function}

As it is seen from the above, the use of other suggested Drude parameters
in the extrapolation of the optical data to lower frequencies leads to
drastically different theoretical predictions for the Casimir pressure.
An interesting suggestion on how to determine $\varepsilon(i\xi)$
using nothing but the optical data in the frequency region where they are
available was proposed \cite{37}. If this were possible,
one could avoid using any extrapolation of the optical data either to
low or high frequencies and remain on the solid grounds of the
measured data. This suggestion is based on the possibility to introduce
a function $f(\omega)$ which is analytic in the upper half-plane
(with possible exclusion of the origin $\omega=0$), which modulus increases
not faster than $|\omega|$ when $|\omega|\to\infty$, and which
suppresses the contribution of frequencies where the optical data are
not measured. It is also assumed that
$f(-\omega^{\ast})=f^{\ast}(\omega)$.
Then, under the assumption that $\varepsilon(\omega)$ is regular or has
at most a first-order pole at $\omega=0$, the Kramers-Kronig relation
takes the form
\begin{eqnarray}
&&
\varepsilon(i\xi)=1+\frac{2}{\pi f(i\xi)}\int_{0}^{\infty}
\frac{\omega\,d\omega}{\omega^2+\xi^2}
\label{eq8} \\
&&
~~\times
\left\{\,
{\rm Im}f(\omega)[{\rm Re}\varepsilon(\omega)-1]\right.
\left.
+{\rm Re}f(\omega)\vphantom{[}{\rm Im}\varepsilon(\omega)
\right\}.
\nonumber
\end{eqnarray}
\noindent
This generalized Kramers-Kronig relation is obtained from Eq.~(\ref{eq3})
by replacing $\varepsilon(\omega)-1$ with
$f(\omega)[\varepsilon(\omega)-1]$.
It is valid for any $\xi$ such that $f(i\xi)\neq 0$. For
$f(\omega)\equiv 1$, Eq.~(\ref{eq8}) coincides with the standard
Eq.~(\ref{eq3}).
Function $f(z)$ was called a {\it window} function \cite{37}.
The following family of window functions was suggested \cite{37}
which could suppress the contribution of frequencies
outside the region where the optical data are measured:
\begin{equation}
f(\omega)=\omega^{2p+1}\left[
\frac{1}{(\omega-\Omega)^{2q+1}}+
\frac{1}{(\omega+\Omega^{\ast})^{2q+1}}\right].
\label{eq9}
\end{equation}
\noindent
Here, $\Omega$ is an arbitrary complex number with
${\rm Im}\Omega<0$,
and $p<q$ are integers. As noted \cite{37}, by taking
sufficiently large values of $p$ one can suppress the contribution
of low frequencies in the integral in Eq.~(\ref{eq8}), where the
optical data are not readily measured, to any desired level.

As an example,  the analytical expression
for the dielectric permittivity of Au along the real frequency axis
was considered \cite{37},
\begin{equation}
\varepsilon(\omega)=1-\frac{\omega_p^2}{\omega(\omega+i\gamma)}+
\sum_{j=1}^{6}
\frac{g_j}{\omega_j^2-\omega^2-i\gamma_j\omega},
\label{eq10}
\end{equation}
\noindent
where the values of the oscillator strengths $g_j$, oscillator
frequencies $\omega_j$ and relaxation parameters $\gamma_j$ were
determined \cite{22} from the fit of
${\rm Im}\varepsilon(\omega)$ to the tabulated optical
data \cite{34}.
Then the values of ${\rm Re}\varepsilon(\omega)$
and ${\rm Im}\varepsilon(\omega)$ from Eq.~(\ref{eq10})
in some restricted frequency region
(wider than in Ref.~\cite{34}) were substituted into Eq.~(\ref{eq8})
with the function $f(\omega)$ defined in Eq.~(\ref{eq9}),
$\Omega=(1-2i)\,$eV, $p=1$ and $q=2$ and 3. It was found that the
obtained $\varepsilon(i\xi)$ is in good agreement with
$\varepsilon(i\xi)$ computed directly from Eq.~(\ref{eq10}) in the
frequency region from 0.16 to 9.7\,eV.

We have attempted to apply Eqs.~(\ref{eq8}) and (\ref{eq9}) to the
immediate optical data \cite{34} using the values $p=1$
and $q=3$ (for this case the best agreement was achieved in
Ref.~\cite{37}). As a result, for $\xi$ from 2.44 to 2.92\,eV
negative values of $\varepsilon(i\xi)$ were obtained.
This could be explained by the proximity of the root of
$f(i\xi)$ at $\xi_0\approx 2.4\,$eV. However, in the
frequency region of $\xi>3\,$eV
the obtained values of $\varepsilon(i\xi)$ differ dramatically from
the values obtained employing the extrapolation of the
optical data \cite{34} by the Drude model either with
most often used or with other suggested parameters.
Moreover, for $\xi>7.8\,$eV [i.e., in the region with no roots of
$f(i\xi)$] $\varepsilon(i\xi)$ once again becomes negative.
One may guess that these anomalies are explained by the fact that
the tabulated optical data \cite{34} are collected from several
different experiments. However, in our opinion
the reason for obtaining such results in
application of Eq.~(\ref{eq8}) to real measured data is the following.
Unlike the standard Kramers-Kronig relation (\ref{eq3}), which uses
only ${\rm Im}\varepsilon(\omega)$, Eq.~(\ref{eq8}) expresses
$\varepsilon(i\xi)$ through both ${\rm Im}\varepsilon(\omega)$ and
${\rm Re}\varepsilon(\omega)$. It should be realized that the quantity
${\rm Re}\varepsilon(\omega)=n^2-k^2$ (where $n$ and $k$ are the real
and imaginary parts of the complex index of refraction) is determined
with much larger error than ${\rm Im}\varepsilon(\omega)=2nk$
(especially in the frequency regions where $n\approx k$).
Because of this, it is preferable to use Eq.~(\ref{eq3}) rather than
Eq.~(\ref{eq8}) when we deal with experimental optical data.
In this regard we stress that the analytical Eq.~(\ref{eq10}) is
in very good agreement with the optical data \cite{34}
for ${\rm Im}\varepsilon(\omega)$. It does not reproduce, however,
the optical data for ${\rm Re}\varepsilon(\omega)$. When we have the
analytic representation for $\varepsilon(\omega)$ (like Eq.~(\ref{eq10})
considered in Ref.~\cite{37}) there is a possibility to select
$\Omega$, $p$ and $q$ in order to have good agreement between
$\varepsilon(i\xi)$ computed from Eq.~(\ref{eq8}) and directly from
Eq.~(\ref{eq10}). If, however, we have only the optical data
for $n$ and $k$ within some
frequency region measured with some errors, this leads to significantly
larger error for ${\rm Re}\varepsilon(\omega)$ than
for ${\rm Im}\varepsilon(\omega)$. Then
 it seems difficult to compute the values of
$\varepsilon(i\xi)$ with sufficiently high precision using
Eq.~(\ref{eq8}).
It should be realized also that Eq.~(\ref{eq8}) is derived under
the assumption that $\varepsilon(\omega)$ is regular or has a
first-order pole at $\omega=0$. Thus, this equation a priori
favours the Drude model which, as argued above, is experimentally
excluded.
Further investigations are needed to determine
whether this elegant method can be used for the comparison of
experiment with theory.

\section{Generalized plasma-like model}

We continue with a discussion of the comparison between the experimental
data \cite{21,22} and the theoretical predictions from using
the Lifshitz formula at zero temperature. Now we combine this formula
with the dielectric permittivity of the generalized plasma-like model
which disregards dissipation properties of conduction electrons but
takes full account of the interband transitions of core electrons
\cite{2,3,22,36}. The generalized plasma-like permittivity is given
by Eq.~(\ref{eq10}) with $\gamma=0$. Along the imaginary frequency axis
it is presented in the form
\begin{equation}
\varepsilon(i\xi)=1+\frac{\omega_p^2}{\xi^2}+
\sum_{j=1}^{6}
\frac{g_j}{\omega_j^2+\xi^2+\gamma_j\xi}.
\label{eq11}
\end{equation}

Numerical computations of the Casimir pressure were performed by the
substitution of Eq.~(\ref{eq11}) into Eqs.~(\ref{eq1}) and (\ref{eq2})
with $\omega_p=8.9\,$eV, as was determined \cite{21,22}.
The obtained differences between the theoretical and mean experimental
Casimir pressures are plotted as dots in Fig.~5. In the same figure,
the borders of a 95\% and 70\% confidence intervals form the solid
and dashed lines, respectively. As can be seen in Fig.~5, all dots lie
inside both confidence intervals. This means that the zero-temperature
Lifshitz theory combined with the generalized plasma-like model
with the most often used value for the plasma frequency is
consistent with the experimental data \cite{21,22}.
The same data were found to be consistent \cite{21,22} with the
theoretical prediction using Eq.~(\ref{eq11}) and the Lifshitz formula
at the laboratory temperature ($T=300\,$K) where the measurements
of the Casimir pressure were performed. This is explained by the fact that
at separations below $1\,\mu$m the plasma-like dielectric permittivity
(\ref{eq11}) leads to negligibly small thermal corrections which are far
below the total experimental error of force measurements. The situation
differs radically when the optical data are extrapolated to low
frequencies by means of the Drude model (\ref{eq4}) or the analytical
Drude-like dielectric permittivity (\ref{eq10}) is used. For such cases
a large thermal correction arises far exceeding the experimental errors
\cite{21,22}. This allowed to experimentally exclude \cite{2,3}
all theoretical
approaches related to the Drude model with either most often used or other
suggested parameters
at a confidence level of 99.9\%.

Now we compare the experimental data \cite{21,22} with the
predictions from the Lifshitz formula at zero temperature
combined with the generalized plasma-like model when other
suggested values for the plasma frequency are used. We have performed
computations of the Casimir pressure by using Eqs.~(\ref{eq1}),
(\ref{eq2}) and (\ref{eq11}) with the largest suggested mean plasma
frequency $\omega_p=8.38\,$eV found \cite{35} (see Sec.~III).
The computational results for $P^{\rm theor}(a)-\bar{P}^{\rm expt}(a)$
are indicated as dots in Fig.~6 within the separation regions (a)
from 160 to 750\,nm and (b) from 350 to 750\,nm.
The solid lines indicate the borders of the 95\% confidence intervals.
For the comparison purposes, the dashed lines in Fig.~6(b) show
the borders of the 70\% confidence intervals. As can be seen in Fig.~6,
the experimental data exclude the theoretical prediction from the
zero-temperature Lifshitz formula with the  value of
$\omega_p^{(5)}$ at a 95\% confidence level within the range of
separations from 160 to 370\,nm. From Fig.~6(b) it follows also that
at a 70\% confidence level the same theoretical predictions are excluded
over a wider separation region from 160 to 480\,nm.
It must be emphasized that for all other
suggested plasma frequencies from
6.82 to 8.38\,eV considered \cite{35}
the magnitudes of computed theoretical Casimir pressures,
$P^{\rm theor}(a)$, are less than for $\omega_p^{(5)}=8.38\,$eV.
As a result, these theoretical predictions are experimentally excluded
at a 95\% confidence level over a wider separation region than in Fig.~6.
The values of the plasma frequency from 8.38\,eV to approximately
8.44\,eV are also excluded at a 95\% confidence level over a bit more
narrow separation region than in Fig.~6. As to the values of the plasma
frequency from 8.45 to 8.65\,eV, they are excluded by the experimental
data \cite{21,22} at a 70\% confidence level over different regions
of separations.

\section{Conclusions and discussion}

In the foregoing we have compared the experimental data of the
experiment on an indirect dynamic measurement of the Casimir
pressure between two parallel Au plates \cite{21,22} with the predictions
of the zero-temperature Lifshitz theory computed
employing the optical data
extrapolated to zero frequency by the Drude model with both
most often used or other suggested
 Drude parameters. We have also performed the comparison of the
same data with the computational results obtained
with the help of the
zero-temperature Lifshitz formula combined with the generalized plasma-like
dielectric permittivity with either most often used or other values
of the plasma frequency.

The main conclusion obtained from these comparisons is that
 the zero-temperature Lifshitz theory combined with
the Drude model is excluded by the experimental data for the
Casimir pressure at short separations below $1\,\mu$m.
In the case when the optical data are extrapolated to low frequencies
by means of the Drude model with most often used parameters, the exclusion
occurs at a 70\% confidence level. If the extrapolation uses
other suggested parameters of the Drude model, the zero-temperature
Lifshitz theory is excluded by the data at a 95\% confidence level.
The theoretical predictions from the zero-temperature Lifshitz formula
combined with the generalized plasma-like dielectric permittivity with
the most often used value of the plasma frequency are shown to be
experimentally consistent. The same theoretical approach but with
other suggested values for the plasma frequency is excluded at a 95\%
confidence level. Keeping in mind that the experimental data for the
Casimir force and Casimir pressure in previous experiments were obtained
at room temperature $T=300\,$K and that the Lifshitz formula at zero
temperature but with room-temperature Drude parameters has no clear
physical meaning, conclusion is made that it is more consistent to
compare all such kind of data with the Lifshitz theory at nonzero
temperature.

The disagreement of the experimental data \cite{21,22} with
theory involving the Drude model with
other suggested parameters \cite{35,42}
at both zero and nonzero temperature (for the latter case it was
demonstrated in Refs.~\cite{2,3,52}) raises several important questions.
The dielectric response of conductors on real electromagnetic field
of sufficiently low frequencies is
described beyond any reasonable doubt by the Drude model. However,
substitution of this model with both
most often used and other suggested sets of
Drude parameters in the Lifshitz formula for the Casimir force at any
temperature (zero or nonzero) results in contradictions with the
experimental data. This suggests that there might be some deep
unclarified differences between fluctuating electromagnetic
field considered in the Lifshitz theory and real electromagnetic
field \cite{27}. Measurements of the optical
properties \cite{35}  with different Au films deposited on different
substrates unequivocally demonstrated that these properties
 depend on the method of preparation of the film and
can vary  from sample to sample.
Observed variations, however, are mostly
determined by the relaxation properties of conduction electrons in thin
films.
The attempt to describe respective optical data by the simple Drude model
with a frequency-independent relaxation parameter results in
sample-to-sample variation of both Drude parameters.
However, one should take proper account of the fact that the Casimir
pressures computed with the help  of the Drude model with {\it any} of
the suggested Drude parameters are experimentally excluded by the
experiments \cite{20,21,22} while the theoretical results
obtained
employing the generalized plasma-like model with the most often used
value of the plasma frequency are experimentally consistent.
Then it is natural to suggest that the dielectric permittivity in the
Lifshitz theory should not be considered in the standard way as
obtained from a response of a metal film to a real electromagnetic field.
It appears as if the dielectric permittivity in the
Lifshitz theory directly accounts for the contribution  of
core electrons, but  treats conduction electrons as a nondissipative
plasma.

As it was mentioned in the Introduction, there are experiments of three
different types with metallic \cite{19,20,21,22,43}, semiconductor
\cite{23,24} and dielectric \cite{25,26} test bodies which cast doubts
on the use of the Drude model in the Lifshitz theory. Some
complicated issues related to all experiments on measuring the Casimir
force were discussed in Sec.~II. Keeping in mind that
experiments on measuring very small forces and separations are rather
complicated, it would be of much interest to have additional independent
confirmation of the obtained results. Such kind of experiments can be
proposed.
For this purpose one should use two micromachined oscillators
with the same Au-coated spheres, as in
Refs.~\cite{20,21,22,43}, but with Au-coatings on the plates made as
suggested in Ref.~\cite{35}. In one oscillator the plate should be coated
with Au following the deposition procedure \cite{35} used
for the sample N1 ($\omega_p^{(1)}=6.82\,$eV and $\gamma^{(1)}=40.5\,$meV).
For the second oscillator the Au coating on the plate should be
performed \cite{35} as for the
sample N5
($\omega_p^{(5)}=8.38\,$eV and $\gamma^{(5)}=37.1\,$meV).
In fact it would be sufficient that the characteristic sizes
of grains in Au-coatings on the two plates be markedly
different \cite{35}.
In this case it is easily seen that the respective difference in the Casimir
pressures computed for the two oscillators
using the Drude model with such different parameters
is several times larger than the total experimental error of
pressure measurements within a wide region of separations. If the mean
Casimir pressures measured with two oscillators would be different,
it would demonstrate the role of relaxation  of conduction
electrons. If, however, in both cases the same Casimir pressures are
obtained, it would confirm that relaxation properties
of conduction electrons do not influence the Casimir effect and
should be
disregarded.
To perform the suggested experiment, the measurement scheme of a
difference-type can be used \cite{53,54}. In this case two halves of
the plate of an oscillator are made using different deposition procedures
(one-half is covered with Au-coating consisting of large grains
and another-half of small grains).
When such a patterned plate moves back and forth below the sphere, the
measured difference Casimir force would be nonvanishing (vanishing)
depending on the role of relaxation of free charge carriers.
Thus, the result of this experiment can give
the ultimate answer to the question whether the relaxation properties
of conduction electrons influence the Casimir force.

%%%%%%%%%%%%%%%%%%%%%%%%%%%%%%%%%%%%%%%%%%%%%%%%%
\section*{Acknowledgments}
The authors are greatly indebted  to R.~S.~Decca for the permission to
use the measurement data of his experiment and for many corrections
and suggestions in the manuscript
which have helped to improve the presentation.
They thank G.~Bimonte for helpful discussion of Sec.~IV.
 G.L.K. and V.M.M. are
grateful to the  Institute
for Theoretical Physics, Leipzig University for kind
hospitality.
They  were supported by Deutsche Forschungsgemeinschaft,
Grant No.~GE\,696/10--1.
%%%%%%%%%%%%%%%%%%%%%%%%%%%%%%%%%%%%%%%%%%%%%%%%%

%%%%%%%%%%%%%%%%%%%
%\end{document}
%%%%%%%%%%%%%%%%%%%%%%%
%%%%%%%%%%%%___FIGURES__%%%%%%%%%%%%%%%%
%%%%%%%%__FIGURE__1__%%%%%%%%%%%%%%%%%%%%%%%%%%%%%%%%%%%%
\begin{figure*}[h]
\vspace*{-4.cm}
\centerline{
\includegraphics{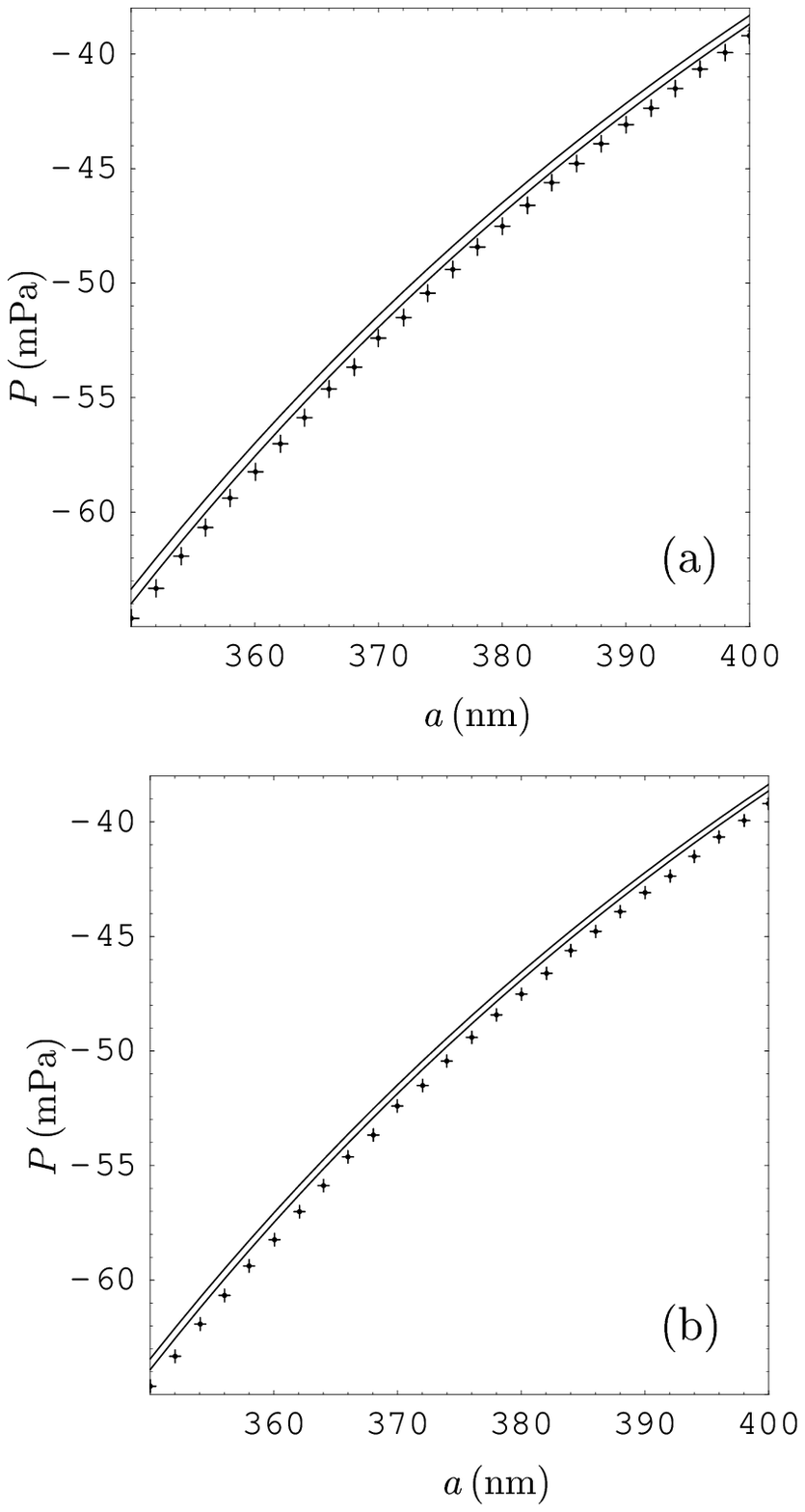}
} \vspace*{-9.cm} \caption{Experimental data for the Casimir
pressure (crosses) as a function of separation and
the theoretical band between the two solid lines computed employing
the Lifshitz formula at $T=0$ and the Drude extrapolation of the
optical data with most often used parameters.
The arms of the crosses and the widths of the bands are determined at
(a) 95\% confidence level and (b) 70\% confidence level.}
\end{figure*}
%%%%%%%%%%%%%%%%%%%%%%%%%%%%%%%%%%%%%%%%%%%%%%%%%%%%%%%%%
%%%%%%%%__FIGURE__2__%%%%%%%%%%%%%%%%%%%%%%%%%%%%%%%%%%%%
\begin{figure*}[h]
\vspace*{-12.cm}
\centerline{
\includegraphics{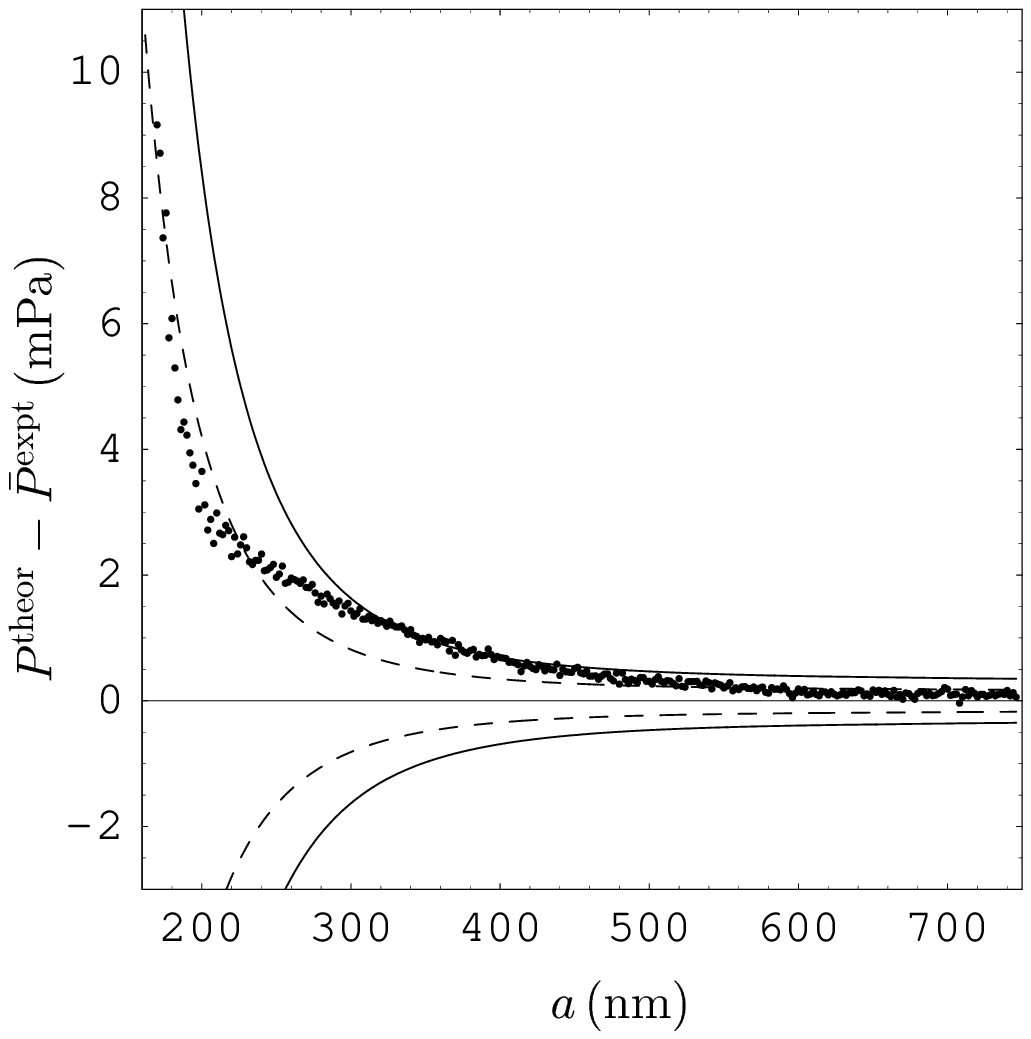}
} \vspace*{-6.5cm} \caption{Differences between theoretical and  mean
experimental Casimir pressures are indicated as dots. Computations are
performed employing
the Lifshitz formula at $T=0$ and the Drude extrapolation of the
optical data with most often used parameters. Solid and dashed lines indicate
the borders of 95\% and 70\% confidence intervals, respectively.}
\end{figure*}
%%%%%%%%%%%%%%%%%%%%%%%%%%%%%%%%%%%%%%%%%%%%%%%%%%%%%%%%%
%%%%%%%%__FIGURE__3__%%%%%%%%%%%%%%%%%%%%%%%%%%%%%%%%%%%%
\begin{figure*}[h]
\vspace*{-4.cm}
\centerline{
\includegraphics{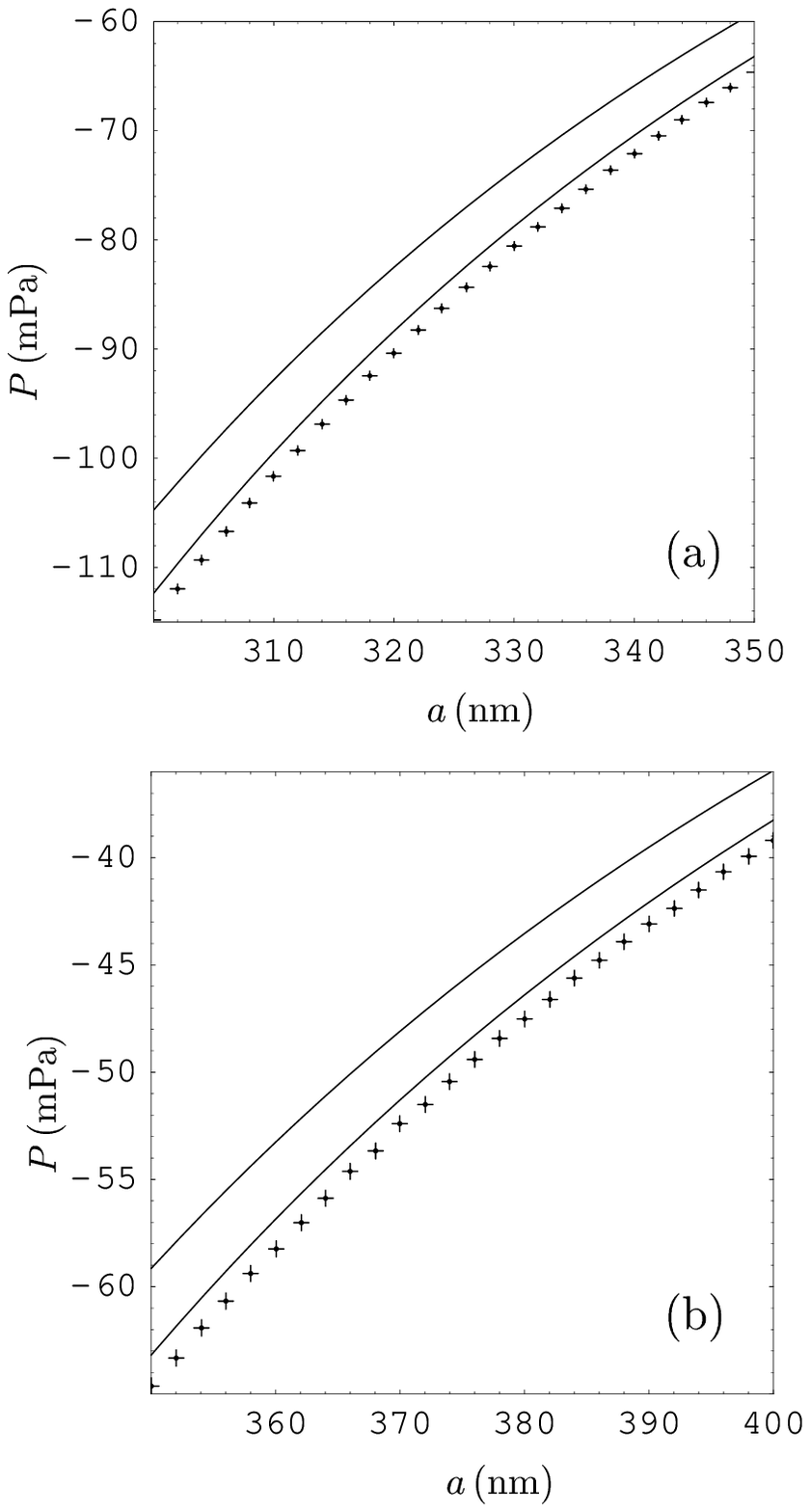}
} \vspace*{-8.5cm} \caption{Experimental data for the Casimir
pressure (crosses) determined at a 95\% confidence level and
the theoretical band between the two solid lines computed employing
the Lifshitz formula at $T=0$ and the Drude extrapolation of the
optical data with different sets of
 parameters for separations
(a) from 300 to 350\,nm and (b) from 350 to 400\,nm.
The widths of the bands are determined at a 95\% confidence level.}
\end{figure*}
%%%%%%%%%%%%%%%%%%%%%%%%%%%%%%%%%%%%%%%%%%%%%%%%%%%%%%%%%
%%%%%%%%__FIGURE__4__%%%%%%%%%%%%%%%%%%%%%%%%%%%%%%%%%%%%
\begin{figure*}[h]
\vspace*{-3.cm}
\centerline{
\includegraphics{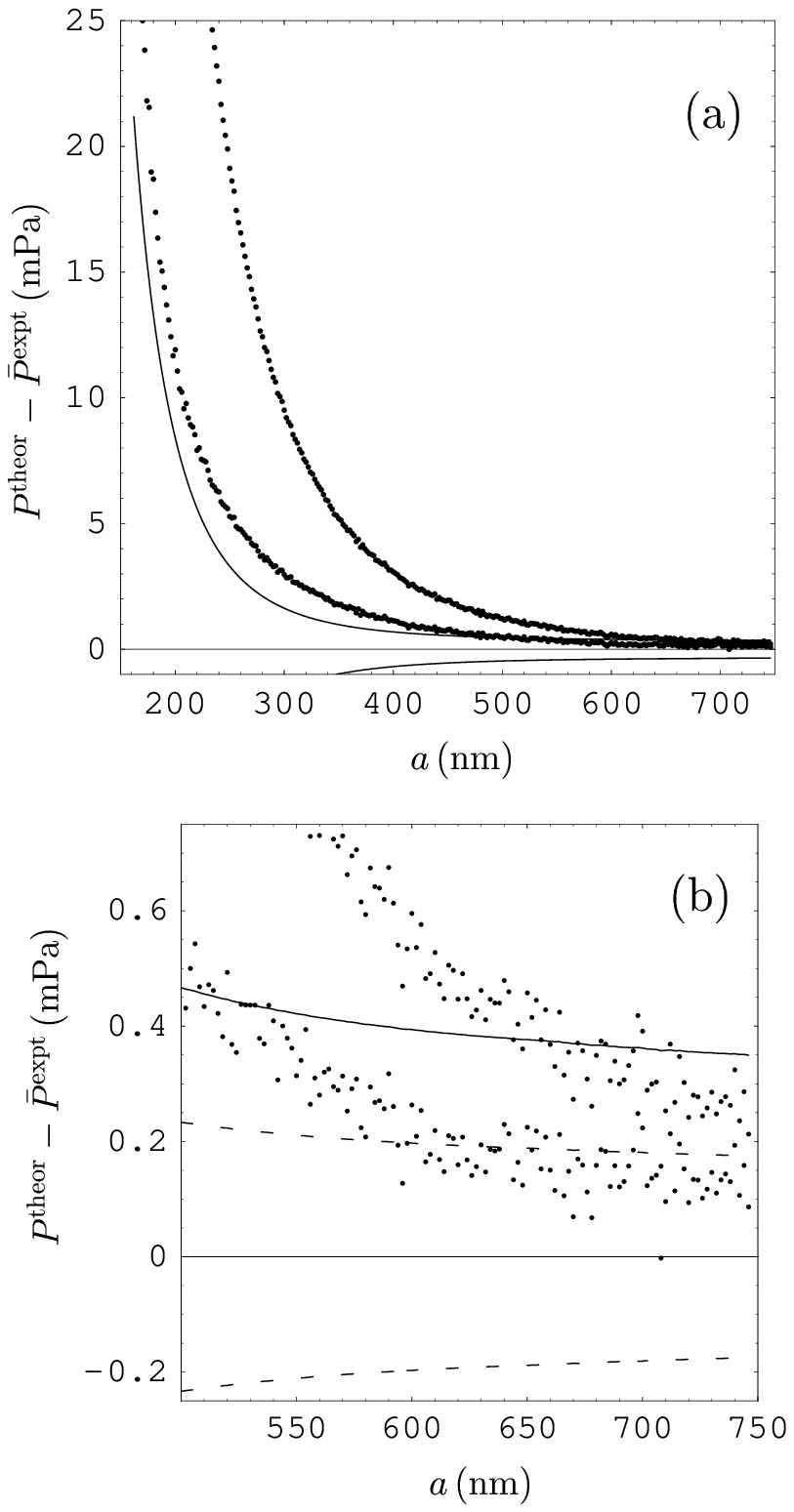}
} \vspace*{-9.5cm} \caption{Differences between theoretical and  mean
experimental Casimir pressures are indicated as dots. Computations are
performed using
the Lifshitz formula at $T=0$ and the Drude extrapolation of the
optical data with different sets of parameters (the upper and lower
sets of dots correspond to smaller and larger plasma frequency,
respectively) for separations (a) from 160 to 750\,nm and
(b) from 500 to 750\,nm. Solid and dashed lines indicate
the borders of 95\% and 70\% confidence intervals, respectively. }
\end{figure*}
%%%%%%%%%%%%%%%%%%%%%%%%%%%%%%%%%%%%%%%%%%%%%%%%%%%%%%%%%
%%%%%%%%__FIGURE__5__%%%%%%%%%%%%%%%%%%%%%%%%%%%%%%%%%%%%
\begin{figure*}[h]
\vspace*{-12.cm}
\centerline{
\includegraphics{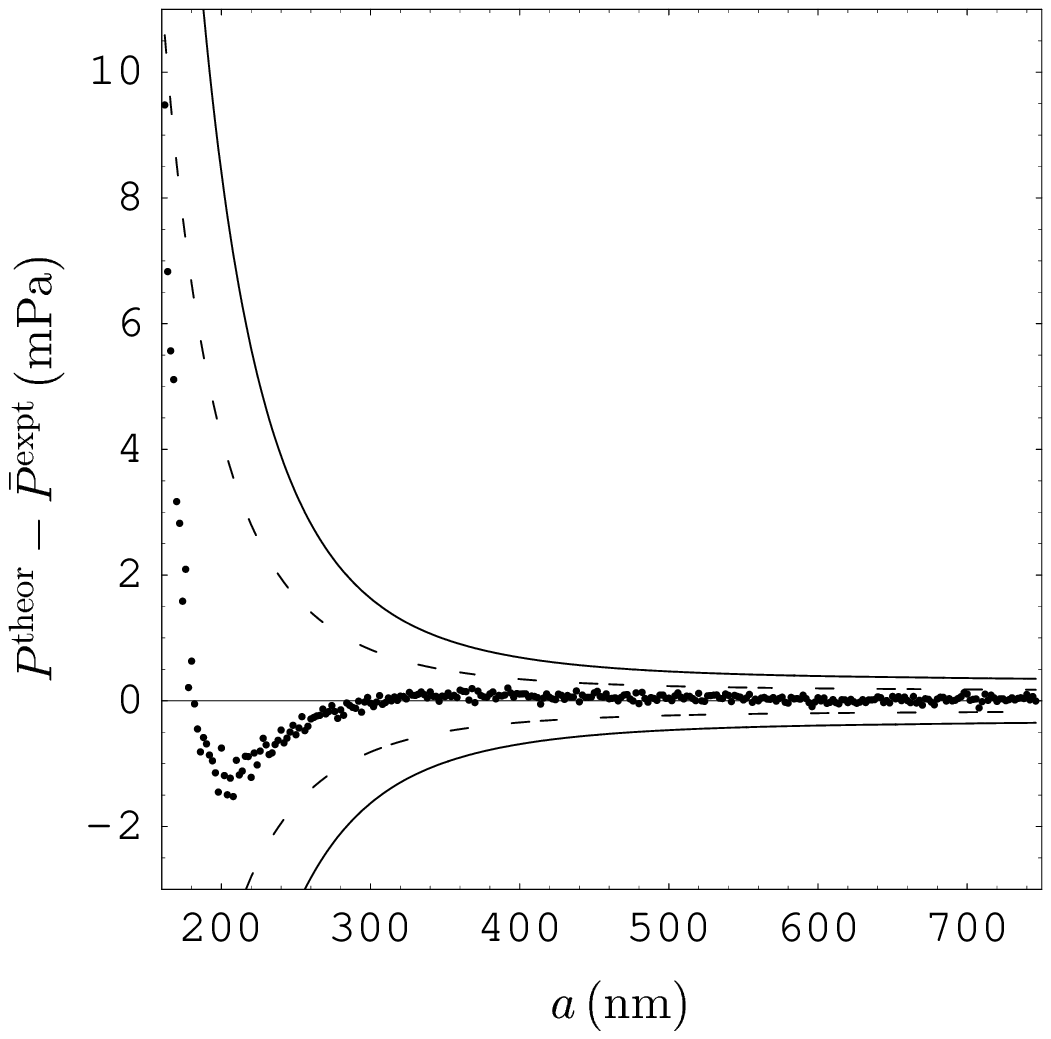}
} \vspace*{-6.5cm} \caption{
Differences between theoretical and  mean
experimental Casimir pressures are indicated as dots. Computations are
performed employing
the Lifshitz formula at $T=0$ and the generalized plasma-like
dielectric permittivity with the most often used value of the
plasma frequency. Solid and dashed lines indicate
the borders of 95\% and 70\% confidence intervals, respectively.}
\end{figure*}
%%%%%%%%%%%%%%%%%%%%%%%%%%%%%%%%%%%%%%%%%%%%%%%%%%%%%%%%%
%%%%%%%%__FIGURE__6__%%%%%%%%%%%%%%%%%%%%%%%%%%%%%%%%%%%%
\begin{figure*}[h]
\vspace*{-3.cm}
\centerline{
\includegraphics{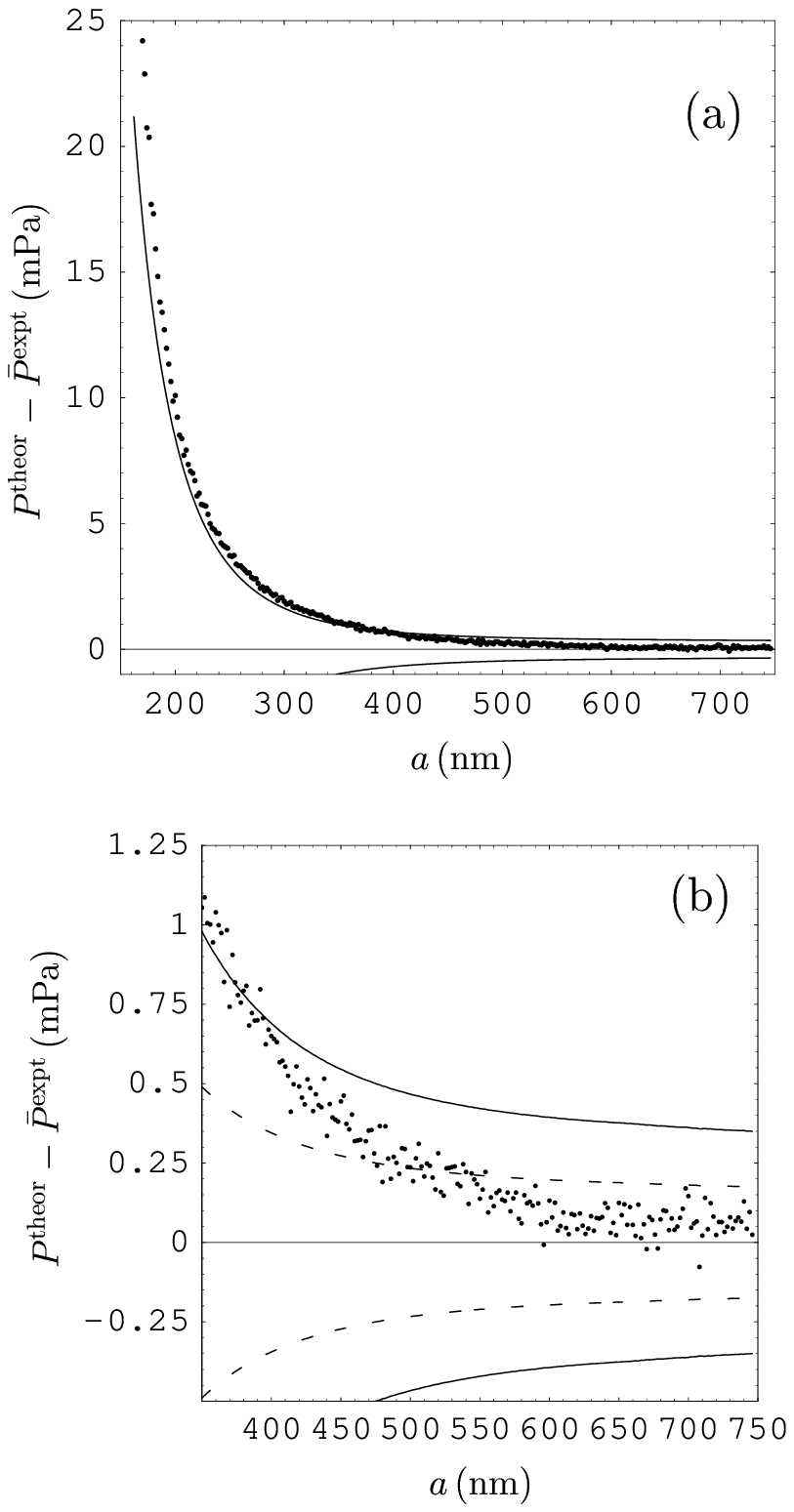}
} \vspace*{-9.5cm} \caption{Differences between theoretical and  mean
experimental Casimir pressures are indicated as dots. Computations are
performed using
the Lifshitz formula at $T=0$ and the generalized plasma-like
dielectric permittivity with  the largest of other suggested
plasma frequencies for separations (a) from 160 to 750\,nm and
(b) from 350 to 750\,nm. Solid and dashed lines indicate
the borders of 95\% and 70\% confidence intervals, respectively.}
\end{figure*}
%%%%%%%%%%%%%%%%%%%%%%%%%%%%%%%%%%%%%%%%%%%%%%%%%%%%%%%%%
%%%%%%%%%%%%%%%%
\end{document}